\documentclass[aps,prb,twocolumn,showpacs]{revtex4-1}
\pdfoutput=1
\usepackage{graphicx}
\usepackage{natbib}
\usepackage[utf8]{inputenc}
\usepackage{epstopdf}
\setcitestyle{super}
\usepackage{color}
\usepackage[normalem]{ulem}
\usepackage{pdfpages}

\newcommand*{\citen}[1]{%
  \begingroup
    \romannumeral-`\x 
    \setcitestyle{numbers}%
    \cite{#1}%
  \endgroup   
}

\makeatletter
\AtBeginDocument{\let\LS@rot\@undefined}
\makeatother

\usepackage{hyphenat}
\begin{document}

\preprint{APS/123-QED}
\title{Insights from experiment and \textit{ab initio} calculations into the glass-like transition in the molecular conductor \mbox{$\kappa$-(BEDT-TTF)$_2$Hg(SCN)$_2$Cl}}
\author{Elena Gati$^{1}$}
\author{Stephen M. Winter$^{1,2}$}
\author{John A. Schlueter$^{3,4}$}
\author{Harald Schubert$^{1}$}
\author{Jens M\"uller$^{1}$}
\author{Michael Lang$^{1}$}
\address{$^{1}$ Physikalisches Institut, J.W. Goethe-Universität Frankfurt(M), SFB/TR49, D-60438 Frankfurt(M), Germany}
\address{$^{2}$ Institut für Theoretische Physik, J.W. Goethe-Universität Frankfurt(M), SFB/TR49, D-60438 Frankfurt(M), Germany}
\address{$^{3}$ Division of Materials Research, National Science Foundation, Arlington, VA 22230, USA}
\address{$^{4}$ Materials Science Division, Argonne National Laboratory, Argonne, IL 60439, USA}
\date{\today}

\begin{abstract}
We present high-resolution measurements of the relative length change as a function of temperature of the organic charge-transfer salt \mbox{$\kappa$-(BEDT-TTF)$_2$Hg(SCN)$_2$Cl}. We identify anomalous features at $T_g\,\approx\,63$\,K which can be assigned to a kinetic glass-like ordering transition. By determining the activation energy $E_A$, this glass-like transition can be related to conformational degrees of freedom of the ethylene endgroups of the organic building block BEDT-TTF. As opposed to other \mbox{$\kappa$-(BEDT-TTF)$_2X$} salts, we identify a peculiar ethylene endgroup ordering in the present material in which only one of the two crystallographically inequivalent ethylene endgroups is subject to glass-like ordering. This experimental finding is fully consistent with our predictions from \textit{ab initio} calculations from which we estimate the energy differences $\Delta E$ and the activation energies $E_A$ between different conformations. The present results indicate that the specific interaction between the ethylene endgroups and the nearby anion layers leads to different energetics of the inequivalent ethylene endgroups, as evidenced by different ratios $E_A/\Delta E$. We infer that the ratio $E_A/\Delta E$ is a suitable parameter to identify the tendency of ethylene endgroups towards glass-like freezing.
\end{abstract}


\maketitle
\section{Introduction}

Organic charge-transfer salts are considered as model systems in the study of the physics of strongly correlated electron systems in low dimensions\cite{Kanoda97,Ishiguro98,Toyota07,Lebed08,Powell11}. They are outstanding by the variety of intriguing ground states, including superconductivity\cite{Lang04,Wosnitza07,Saito11}, Mott insulating\cite{Kanoda97,Kagawa05,Gati17}, multiferroic\cite{Lunkenheimer12,Lang14} and spin-liquid states\cite{Shimizu03,Itou08}. This diversity reflects a high tunability of interaction strength and frustration which can be accessed in laboratory environments by the application of moderate pressures or slight chemical modifications\cite{Kanoda97,Lefebvre00,Kawamoto97}. Besides that, organic charge-transfer salts are available in very clean single-crystalline form, as evidenced by the observation of quantum oscillations in low fields\cite{Wosnitza96,Kartsovnik08}. Nevertheless, these systems tend to be susceptible to intrinsic disorder which arises from structural degrees of freedom of the large-sized molecular building blocks\cite{Pouget96,Ravy88}. This type of intrinsic disorder occurs whenever a certain structural unit can adopt two different orientations which are almost degenerate in energy. In this situation, the structural elements often  cannot order for kinetic reasons and instead, tend to undergo glass-like transitions around a characteristic temperature $T_g$. The properties of this ``glass-like'' state in otherwise well-ordered solids are found to be similar\cite{Mueller11,Gugenberger92} to conventional glass formers\cite{Debenedetti01}, i.e., undercooled liquids, such as glucose. The term glass implies that the relaxation of the structural degrees of freedom becomes so slow that thermal equilibrium cannot be reached and short-range structural order with residual disorder is ``frozen''. An important characteristic of this glass-like behavior is that the conditions of the non-equilibrium state depend on the cooling rate\cite{Hartmann14}. Conversely, the amount of frozen disorder can be very well controlled in a reversible way by varying the cooling rate. Thus, the organic charge-transfer salts have been identified as a suitable test ground to investigate the delicate interplay between strong electronic correlations and disorder effects\cite{Sasaki12,Mueller15,Analytis06,Sasaki08,Sano10,Diehl15}, a problem that is relevant for any solid-state realization of a strongly correlated electron system and treated in theoretical models, such as the Mott-Anderson model\cite{Shinaoka09,Shinaoka09b,Byczuk05}.

Particularly well-suited materials for these investigations are members of the widely studied family of \mbox{$\kappa$-(BEDT-TTF)$_2X$}  charge-transfer salts ($X$ represents a monovalent anion, see Fig.\,\ref{fig:structure}) for the following reasons. First, this material class reflects the rich phenomenology associated with strong correlations\cite{Kanoda97,Toyota07}, as represented by the multiferroic Mott insulator\cite{Kanoda97,Lunkenheimer12,Miyagawa95} $X$\,=\,Cu[N(CN)$_2$]Cl, the superconductors\cite{Lang04} $X\,=\,$Cu[N(CN)$_2$]Br and Cu(SCN)$_2$ and the spin-liquid candidate system\cite{Shimizu03} $X$\,=\,Cu$_2$(CN)$_3$. Second, structural degrees of freedom, susceptible to glass-like ordering, are inherent to the molecular building block BEDT-TTF which stands for C$_6$S$_8$[C$_2$H$_4$]$_2$ (bis-ethylenedithio-tetrathiafulvalene). Here, the ethylene endgroups [C$_2$H$_4$] (abbreviated as EEG hereafter) can adopt two different relative orientations (see Fig. \ref{fig:structure} (c)), either an eclipsed (\textit{E}) or a staggered (\textit{S}) one when viewed along the central C$=$C bond. The population of these two conformations is often thermally disordered at room temperature, with a tendency to adopt one of the configurations upon lowering the temperature $T$ (see Refs.\,\citen{Hartmann14, Mueller15} for an overview). Whereas the salt $X\,=\,$Cu$_2$(CN)$_3$ does not reveal any signatures of a glass-like ordering\cite{Manna10,Jeschke12} of the EEGs at low $T$, despite thermal disorder at room temperature, the three other salts mentioned above all undergo glass-like transitions\cite{Saito99,Akutsu00,Mueller02} around $T_g\,\approx\,70$\,K. Various experiments demonstrated that the dynamics of the EEGs \cite{Kuwata11} strongly affect the ground-state properties of these salts. In particular, the properties of the two superconducting salts $X\,=\,$Cu[N(CN)$_2$]Br and Cu(SCN)$_2$, including their critical temperature\cite{Analytis06,Su98,Stalcup99,Taylor08} $T_c$, were found to be strongly sensitive on the cooling procedure through $T_g$. In case of the former salt $X\,=\,$Cu[N(CN)$_2$]Br, it was even possible to reversibly tune the system from a metal to a Mott insulator\cite{Hartmann15,Mueller17} by increasing the cooling rate $|q|$. Recently, it was argued that this large effect cannot solely be attributed to the effects of disorder, but also to changes of the interaction strength\cite{Guterding15,Mueller17} which result from different hopping energies in the eclipsed and staggered configuration.

This strong influence of the orientational degrees of freedom of the EEGs on the physical properties motivates the interest in understanding and modeling the EEG behavior. It is likely and also found experimentally that the details of the EEG ordering depend on the specific system. This concerns not only the preferred orientation at low $T$, but also the occurrence of a glass-like transition: As indicated by the examples given above, only a few, but not all \mbox{$\kappa$-(BEDT-TTF)$_2X$} salts undergo a glass-like transition. In a comparative experimental and theoretical \textit{ab initio} study by Müller \textit{et al.} \cite{Mueller15} on a wide range of \mbox{$\kappa$-(BEDT-TTF)$_2X$} salts, it was argued that the distinct behavior can be explained consistently by considerations of the energetics of the different conformations. Importantly, the energies are mostly determined by the specific anion-EEG interaction. This emphasizes that not only EEG degrees of freedom are involved in the glass-like ordering process, but also their coupling to the anions \cite{Wolter07} leading to a collective motion of EEG and anion molecules that freezes at $T_g$ \cite{Mueller15}. For the first time, these calculations allow to make predictions on the occurrence of a glassy EEG state in the \mbox{$\kappa$-(BEDT-TTF)$_2X$} salts. 

In this paper, we apply these theoretical methods to the charge-ordered\cite{Drichko14} ferroelectric\cite{Gati17b} salt \mbox{$\kappa$-(BEDT-TTF)$_2$Hg(SCN)$_2$Cl} and identify it as a special candidate for understanding the mechanisms of glass-like EEG ordering. We predict only half of the EEGs undergo a glass-like transition, whereas the other half is predicted to order smoothly. Such a distinct behavior in one single system was not resolved yet for any other \mbox{$\kappa$-(BEDT-TTF)$_2X$} salt. In order to verify this theoretical prediction for \mbox{$\kappa$-(BEDT-TTF)$_2$Hg(SCN)$_2$Cl}, we conduct measurements of the thermal expansion coefficient on this salt to unravel the potential glass-like EEG ordering. This technique was successfully used in the past to identify glass-like ordering in \mbox{$\kappa$-phase} organic charge-transfer salts\cite{Mueller02}. The high sensitivity of this method is due to its inherent sensitivity to structural variations.

\section{Methods}

Single crystals of \mbox{$\kappa$-(BEDT-TTF)$_2$Hg(SCN)$_2$Cl} were synthesized by the standard electrocrystallization technique. We followed the strategy reported in Ref.\,\citen{Konovalikhin92}, however with minor modifications of the synthesis route, as follows. Pure TCE (1,1,2-Trichloroethane) was employed as a solvent with a mixture of Hg(SCN)$_2$ and PPNCl (bis(triphenylphosphoranylidene)ammonium chloride) in a molar ratio of 1:1 serving as the electrolytes. The electrolyte was given in a ten-fold excess to the solution in relation to the BEDT-TTF. This results in a typical composition of 75 mg of BEDT-TTF, 642 mg of Hg(SCN)$_2$ and 1132 mg of PPNCl in 100 ml solvent. A constant current of 0.2 $\mu$A was applied to platinum electrodes, resulting in a voltage of 0.1\,V – 0.3\,V. Crystal growth was performed at a temperature of 20$^\circ$C and crystals were collected after 4-5 weeks. Crystals were characterized by means of resistance measurements in order to identify the characteristic metal-insulator transition\cite{Yasin12} in this compound at $T_{MI}\,\approx\,30$\,K. For the thermal expansion measurements, the crystals were oriented by eye resulting in a maximum misalignment of 5$^\circ$. \\
Measurements of the relative length change $\Delta L_i(T)/L_i$ ($i\,=\,a,b,c$), with $\Delta L_i(T) = L_i(T)-L_i(T_0)$ and $T_0$ a reference temperature (here chosen to be $T_0\,=\,200\,$K), were performed by using a home-built capacitive dilatometer. The design of this dilatometer is similar to the one described in Ref.~\citen{Pott83} and reaches a maximum sensitivity of $\Delta L/L \,\approx\,10^{-10}$. The thermal expansion coefficient 
\begin{eqnarray}
\alpha_i(T) &=& \frac{1}{L_i}\frac{\textnormal{d}L_i}{\textnormal{d}T} \\ &\approx & \frac{1}{L_i(300\,\textnormal{K})} \frac{\Delta L_{i}(T_2)-\Delta L_{i}(T_1)}{T_2-T_1} \\ \textnormal{with}\,T&=&(T_1+T_2)/2\,\textnormal{and}\,i\,=\,a,b,c
\end{eqnarray}
was calculated numerically from the $\Delta L_i(T)/L_i$ data using the following procedure: The $\Delta L_i(T)/L_i$ data were divided into equidistant intervals of typically $\Delta T\,=\,0.3\,$K. In each of these intervals the mean slope was determined from a linear regression. The mean slope together with the mean temperature in this interval correspond to one data point in the $\alpha_i$ vs. $T$ representation. 
Measurements of $\Delta L_i(T)/L$ (and correspondingly $\alpha_i(T)$) were performed upon warming and cooling in the temperature range $5\,\textnormal{K}\,\le\,T\,\le\,200\,\textnormal{K}$. The temperature was controlled by a LakeShore 340 controller using a heating rate of $q_h\,=\,+1.5\,$K/h and cooling rates ranging between $(1.20\,\pm\,0.05)\,\textnormal{K/h}\,\le\,|q_c|\,\le\,(20.7\,\pm\,0.3)\,$K/h.

 \textit{Ab initio} calculations were performed using ORCA\cite{Neese12} at the B3LYP/def2-SV(P) level, based on the structural data at $T\,=\,300$\,K reported in Ref.~\citen{Drichko14}, and following the procedure from Ref.~\citen{Mueller15}. In particular, the positions of the EEGs were relaxed, while the remaining atoms were fixed. We stress that temperature-induced thermal expansion effects do not lead to significant changes in the resulting energy scheme. The interaction between the dimer and anion layer were approximated via the OPLS-aa forcefield\cite{Jorgensen88}, with Lennard–Jones parameters from the GROMACS set\cite{Pronk13}. The anion charge distribution was estimated from Mulliken analysis of B3LYP/def2-SV(P) calculations on small fragments. For transition state calculations, ORCA's hessian mode following algorithm was employed.

\section{Theory: \textit{Ab initio} calculations}

\begin{figure}
\includegraphics[width=\columnwidth]{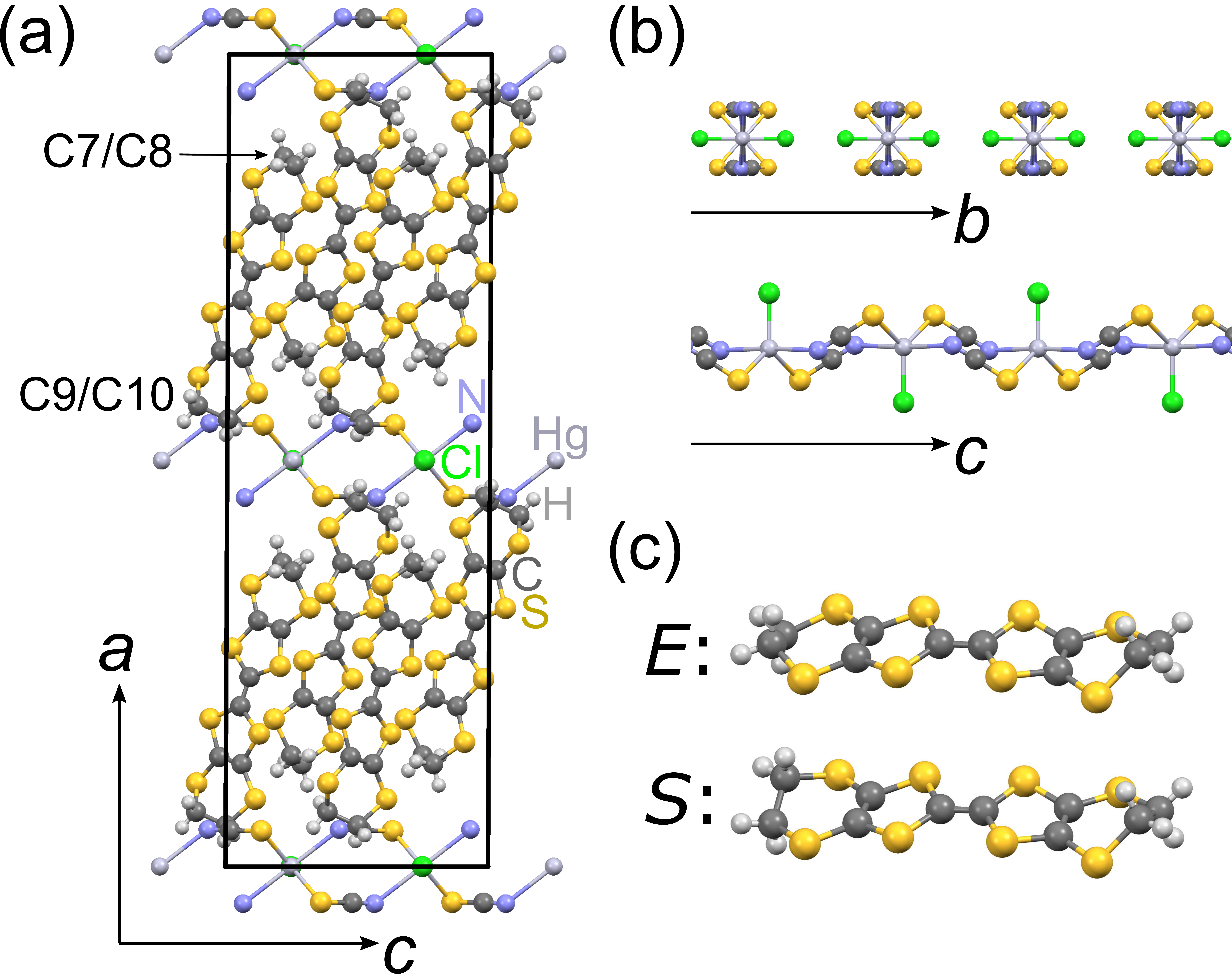} 
\caption{Structure of \mbox{$\kappa$-(BEDT-TTF)$_2$Hg(SCN)$_2$Cl}: (a) Alternating layers of (BEDT-TTF)$_2^+$ and (Hg(SCN)$_2$Cl)$^-$ along the out-of-plane $a$ axis. C7/C8 and C9/C10 refer to the carbon atoms of the two crystallographically inequivalent ethylene endgroups in the present material which we label with inner and outer EEGs, respectively; (b) Polymeric anion layer (Hg(SCN)$_2$Cl)$^-$ viewed along the in-plane $b$ and $c$ axes, indicating the dominant chain-like character along the $c$ axis; (c) Orientational degrees of freedom of the ethylene endgroups (EEGs) of the BEDT-TTF molecule. \textit{E} refers to the eclipsed configuration, \textit{S} to the staggered configuration.}
\label{fig:structure}
\end{figure}

\begin{figure}[h!]
\includegraphics[width=\columnwidth]{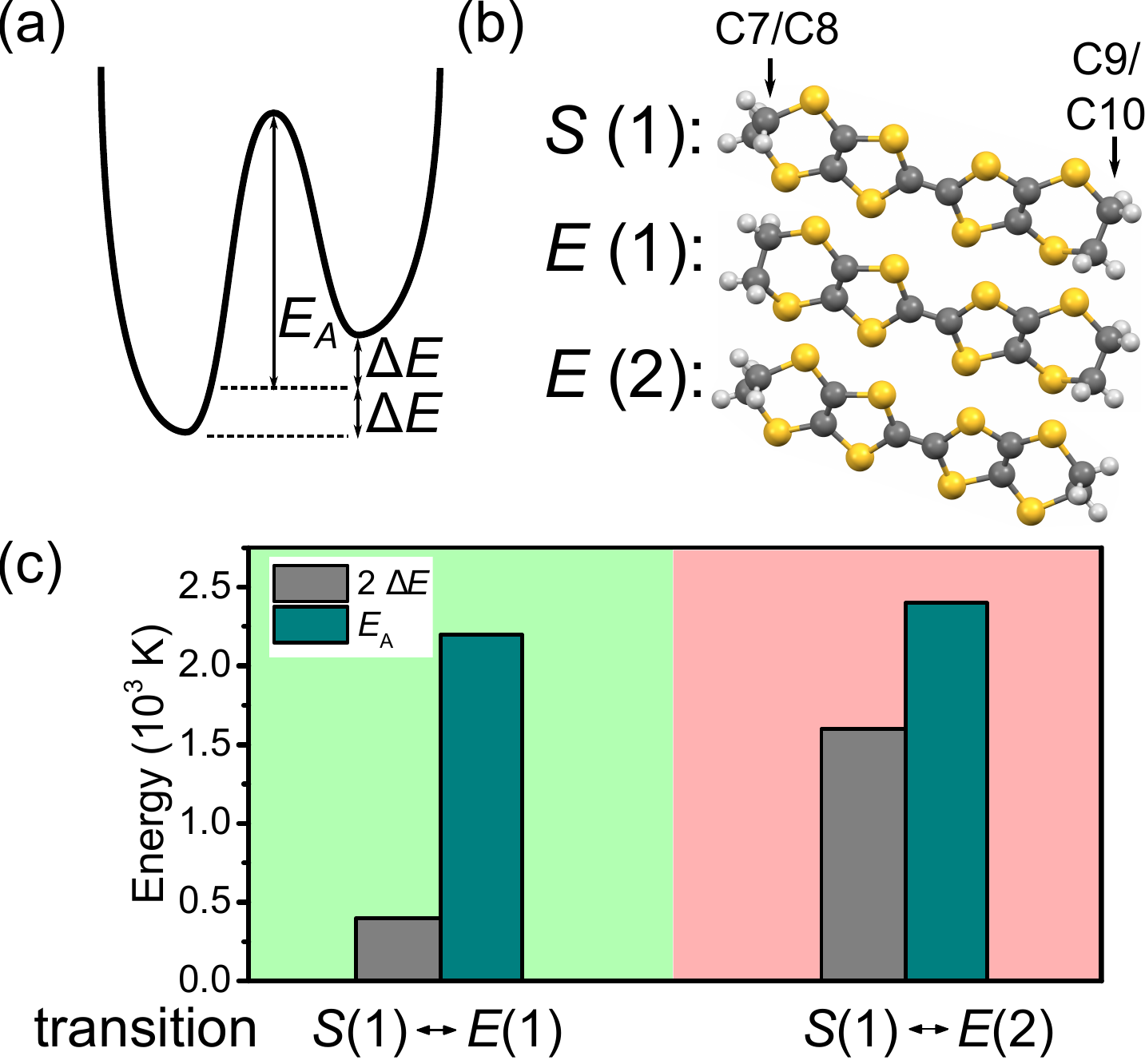} 
\caption{(a) Schematic two-level system in which two states with energy difference $2\Delta E$ are separated by an activation barrier $E_A$; (b) View on BEDT-TTF molecules in three different ethylene endgroup (EEG) conformations, which can be realized in \mbox{$\kappa$-(BEDT-TTF)$_2$Hg(SCN)$_2$Cl}. The majority conformation is a staggered conformation, labeled as \textit{S}(1). The configuration \textit{E}(1) (\textit{E}(2)) is obtained from \textit{S}(1) by changing the orientation of the inner EEGs containing carbon atoms C7 and C8 (outer EEGs containing carbon atoms C9 and C10); (c) Computed energy values $2\Delta E$ and $E_A$ for the processes \textit{S}(1)$\,\leftrightarrow\,$\textit{E}(1) and \textit{S}(1)$\,\leftrightarrow\,$\textit{E}(2) in \mbox{$\kappa$-(BEDT-TTF)$_2$Hg(SCN)$_2$Cl}. The green background color indicates that our calculations predict a glass-like transition for this process, the red background color indicates that no glass-like transition is predicted.} 
\label{fig:conformations}
\end{figure}

A two-level model is often used to describe glass-like transitions\cite{Hartmann14} (see Fig. \ref{fig:conformations} (a)). In this model, two states with energy difference $2\Delta E$ are separated by an activation barrier of size $E_A$ leading to a thermally-activated relaxation time $\tau\,\propto\,\exp(E_A/(k_B T))$. In the following, we discuss possible conformations of the EEGs in \mbox{$\kappa$-(BEDT-TTF)$_2$Hg(SCN)$_2$Cl} and our computational results on the energy scheme in a simple two-level model, including $2\Delta E$ and $E_A$. We distinguish the inequivalent EEGs within each dimer by their distance to the anion layer (see Fig. \ref{fig:structure} (a)): The outer EEGs (containing carbon atoms C9 and C10) almost penetrate the anion layer, whereas the inner ones (carbon atoms C7 and C8) are shifted away from the anion layer. In contrast to \mbox{$\kappa$-(BEDT-TTF)$_2$Cu[N(CN)$_2$]$Z$} ($Z\,=\,$Br, Cl), both the inner and outer EEGs are disordered\cite{Drichko14} at room temperature in \mbox{$\kappa$-(BEDT-TTF)$_2$Hg(SCN)$_2$Cl}. This gives rise to four possible EEG conformations within each molecule in the latter compound (sixteen total conformations for each dimer). The majority conformation at room temperature sees both molecules in a staggered orientation that we denote as \textit{S}(1), shown in Fig.\,\ref{fig:structure} (a). The other three molecular conformations are labeled with \textit{E}(1), \textit{E}(2) and \textit{S}(2) and are obtained by changing the orientation of the inner (\textit{E}(1)), the outer (\textit{E}(2); Fig. \ref{fig:conformations} (b)) or both EEGs (\textit{S}(2)). The sixteen total conformations of each dimer are then obtained as different combinations of these four molecular conformations. According to recent structural investigations\cite{Drichko14}, the inner EEGs are significantly more disordered at room temperature than the outer EEGs: This results in distinctly higher occupancies of the molecular conformations \textit{S}(1) and \textit{E}(1) of 43\% and 35\% ($>$\,25\%), respectively, compared to occupancies of 12\% for \textit{E}(2) and 10\% for \textit{S}(2) ($<$\,25\%).

In order to explore the relative stability of the various conformations, we estimated the energies of the various conformations of \mbox{$\kappa$-(BEDT-TTF)$_2$Hg(SCN)$_2$Cl} via {\it ab-initio} calculations as outlined in section II.  In agreement with the results of structure determination\cite{Drichko14}, our calculations yield that \textit{S}(1) is the most stable molecular conformation. From this starting point, we considered the energy cost $E_i$ to convert one molecule from \textit{S}(1) to the \textit{E}(1), \textit{E}(2), and \textit{S}(2) conformation. After averaging over all dimer conformations according to the experimental room temperature occupancies (see Supplementary Information for details \cite{Supplement}), we find energy differences $2\Delta E_i$ (see Fig. \ref{fig:conformations} (c)):

\begin{eqnarray*}
2 \Delta E_{E(1)} &=& |E_{E(1)}-E_{S(1)}| = (400 \pm 100)\,\textnormal{K}, \\
2 \Delta E_{E(2)} &=& |E_{E(2)}-E_{S(1)}| = (1600 \pm 200)\,\textnormal{K}, \\
2 \Delta E_{S(2)} &=& |E_{S(2)}-E_{S(1)}| = (1900 \pm 200) \,\textnormal{K,}
\end{eqnarray*}

where the variance indicates the weighted standard deviation of the computed values.
These relative energies are in line with the experimentally determined occupancies of the different conformations; those conformations with lower experimental room temperature occupancy are found to have higher energies.

\begin{figure}
\includegraphics[width=\columnwidth]{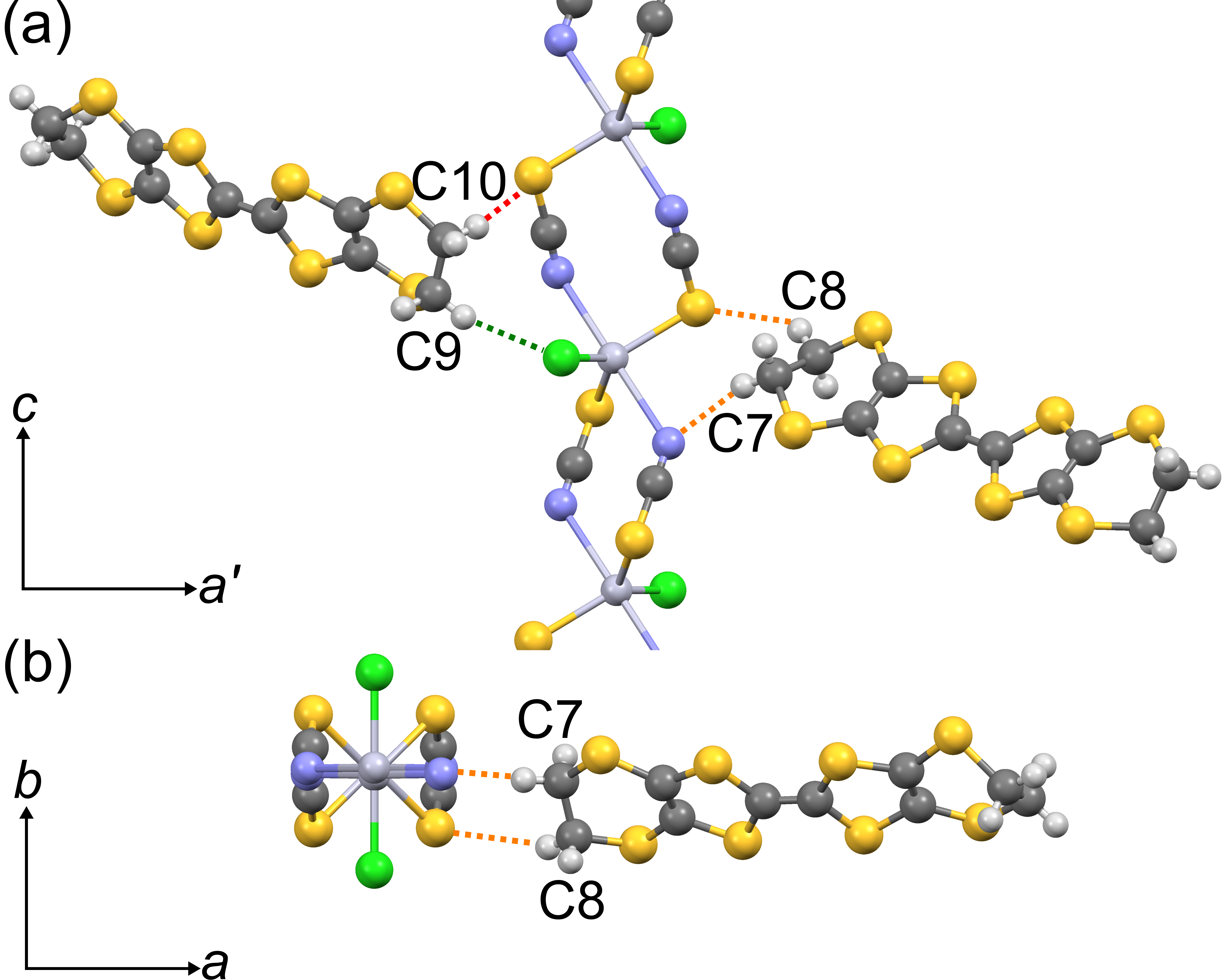} 
\caption{Preferred ethylene endgroup (EEG) conformation (staggered \textit{S}(1)) relative to the nearby anion layer for \mbox{$\kappa$-(BEDT-TTF)$_2$Hg(SCN)$_2$Cl}, viewed within the $a^\prime c$ plane (a) and the $ab$ plane (b). $a^\prime$ indicates a small rotation of the $a$ axis around the $c$ axis. Green and brown dashed lines indicate close contacts of the outer EEGs (labeled with C9 and C10) to the Cl and (SCN) ligands in the nearby anion layer. Orange lines indicate close contacts of the inner EEGs (labeled with C7 and C8) to the (SCN) ligands in the nearby anion layer.}
\label{fig:anion-coupling}
\end{figure}

The relative stability of the different conformations is strongly influenced by the interactions of the EEGs with the nearby anion layer. In \mbox{$\kappa$-(BEDT-TTF)$_2$Hg(SCN)$_2$Cl}, the non-planar polymeric anions (see Fig. \ref{fig:structure} (b)) form a chain-like structure that consists of Hg$^{2+}$ coordination polymers with bridging (SCN)$^-$ ligands and short-side chains formed by the terminal ligand Cl$^-$ coordinated to each Hg$^{2+}$. The two inequivalent EEGs of each BEDT-TTF molecule are embedded in a distinct local environment leading to different coupling paths between the EEGs and the anions (see Fig. \ref{fig:anion-coupling} (a)): The inner EEGs (with carbon atoms C7 and C8) possess only close (SCN)$\cdots\,$H contacts. In contrast, the outer EEGs couple to the anion via (SCN)$\cdots\,$H and Cl$\cdots\,$H contacts. The large $2\Delta E_i$ values for $i\,=\,E(2)\,\textnormal{and}\,S(2)$ which both involve a change of the orientation of the outer EEGs suggest that this EEG conformation is rigidly confined by the coupling to the anion layer. In contrast, the (SCN)$\cdots\,$H interaction to the inner EEGs seems to be rather ineffective in energetically distinguishing the two orientations, as represented by a comparably small $2\,\Delta E_{E(1)}\,=\,330$\,K. Thus, the inner EEGs are closer to a metastable state than the outer EEGs.

From transition state calculations, we have also estimate activation energies (see Fig. \ref{fig:conformations} (c)) for the transition from \textit{S}(1) to \textit{E}(1) to be $E_{A,1}\,\approx\,2200\,$K and for the transition from \textit{S}(1) to \textit{E}(2) to be $E_{A,2}\,\approx\,2400\,$K. We note that we omit a discussion of the process \textit{S}(1)\,$\leftrightarrow$\,\textit{S}(2), as this transition involves changing the orientation of both EEGs. Assuming now that the behavior of the two EEGs in one BEDT-TTF molecule can be considered independently, this implies that the process \textit{S}(1)\,$\leftrightarrow$\,\textit{S}(2) can be decomposed into a two-step transition via an intermediate eclipsed conformation. Thus, the behavior of this process is either dominated by the process \textit{S}(1)\,$\leftrightarrow$\,\textit{E}(1) or \textit{S}(1)\,$\leftrightarrow$\,\textit{E}(2). The given values for $E_{A,i}$ are in the range of $E_A$ values obtained for other \mbox{$\kappa$-(BEDT-TTF)$_2$X} salts (2000\,K$\,\le\,E_A\,\le\,$3100\,\,K), determined by various thermodynamic and transport experiments\cite{Mueller02,Akutsu00,Miyagawa95,Wzietek96} as well as by \textit{ab initio} calculations\cite{Mueller15}. In Ref.~\citen{Mueller15} it was argued that glass-like freezing occurs whenever two confirmations are similar in energy compared to the activation energy of their interchange, i.e., whenever the ratio $E_A/\Delta E$ is sufficiently large. Based on a comparative study of the computed $E_A/\Delta E$ ratios for various \mbox{$\kappa$-phase} BEDT-TTF salts, a threshold value\cite{Mueller15} of $E_A/\Delta E\,\gtrsim\,5$ was found empirically. For \mbox{$\kappa$-(BEDT-TTF)$_2$Hg(SCN)$_2$Cl}, we find $E_{A,1}/\Delta E_{E(1)}\,\sim\,11$ for the inner EEGs and $E_{A,2}/\Delta E_{E(2)}\,\sim\,3$ for the outer EEGs (see Fig. \ref{fig:conformations} (c)). We note that $E_{A,1}/\Delta E_{E(1)}$ is the largest ratio among all investigated \mbox{$\kappa$-(BEDT-TTF)$_2X$} salts\cite{Mueller15} emphasizing the large metastability of this conformation.  Applying the empirical threshold proposed in Ref.~\citen{Mueller15}, the calculations predict only one glass-like transition in the inner EEGs -- despite the existence of two inequivalent EEGs with potential for glassy freezing. The outer EEGs are expected to order smoothly as for $X$ = Cu$_2$(CN)$_3$.

\section{Experiment: Thermal expansion measurements}

\subsection{Phenomenology of glass-like transitions in thermal expansion measurements}

Before discussing the salient results of our thermal expansion studies on the organic charge-transfer salt \mbox{$\kappa$-(BEDT-TTF)$_2$Hg(SCN)$_2$Cl}, we first introduce the main signatures of glass-like transitions in thermodynamic quantities (see also Ref.\,\citen{Mueller02} for a detailed discussion). In order to study thermal equilibrium properties, the experimental observation time $\Delta t$ has to be much larger than the relaxation time of the system $\tau$, i.e., $\Delta t\,\gg\,\tau$. If this criterion is violated, non-equilibrium phenomenon are observed. This is the case for glass (and glass-like) transitions in which the relaxation time of a system slows down so dramatically with lowering $T$ that the relaxation time $\tau$ can reach the value of the observation time $\Delta t$.  Then equilibrium conditions cannot be reached anymore and a glass transition into a metastable state with short-range order occurs at $T_g$. The long relaxation time of the system beyond experimental observation times affects thermodynamic properties in the following ways:

(1) As the associated motion is frozen on experimental timescales below $T_g$, it does not contribute to thermodynamic quantities, such as the specific heat $C(T)$ or the thermal expansion coefficient $\alpha(T)$. In contrast, above $T_g$ these degrees of freedom can be thermally excited and therefore contribute additionally to $C(T)$ and $\alpha(T)$. As a consequence, a step-like increase of $C(T)$ and step-like changes\cite{fussnote1} of $\alpha(T)$ at $T_g$ are expected upon warming. 

(2) In contrast, the cooling behavior of a glass-forming system is expected to be distinctly different from the warming behavior resulting in a strong thermal hysteresis. Whereas upon cooling a smooth anomaly is expected, the discontinuous step-like feature upon warming is usually accompanied by characteristic over- and undershoots. This behavior reflects the strong tendency of a system to achieve an equilibrium state as soon as $T_g$ is approached from below. 

(3) Last, the behavior of the system is strongly dependent on the cooling rate. In particular, the glass transition temperature $T_g$ is strongly affected. On the one hand, the faster a system is cooled, the smaller is the experimental observation time $\Delta t$. On the other hand, the relaxation time increases with lowering $T$. Correspondingly, the criterion $\Delta t\,\approx\,\tau$ is fulfilled at higher temperatures and $T_g$ increases with cooling rate $|q|$. Given the definition of $q\,=\,\Delta T/\Delta t$, the glass transition temperature can be defined based on the criterion $\Delta t\,\approx\,\tau$ as follows \cite{Gugenberger92,Nagel00},
\begin{equation}
-|q|\,\cdot\,\frac{\textnormal{d}T}{\textnormal{d}\tau}\bigg\vert_{T_g}\,\simeq\,1.
\end{equation}
The three main characteristics listed above clearly discriminate a glass transition from a thermodynamic phase transition which takes place in equilibrium.

\subsection{Results}

\begin{figure}
\includegraphics[width=\columnwidth]{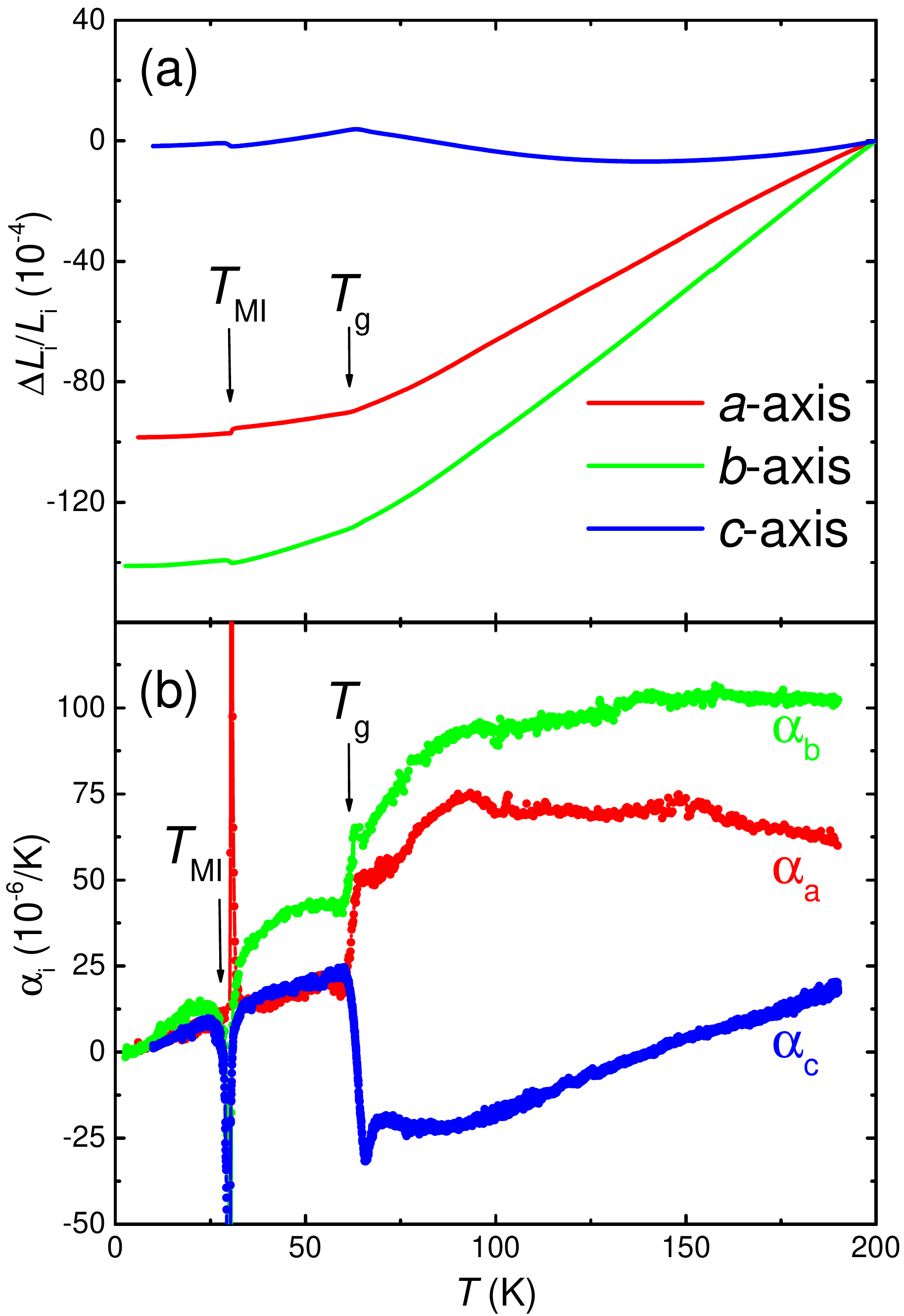} 
\caption{Relative length change $\Delta L_i(T)/L_i$ normalized to the reference temperature $T_0\,=\,200$\,K (a) and thermal expansion coefficient $\alpha_i(T)\,=\,L_i^{-1}\,\textnormal{d}L_i/\textnormal{d}T$ (b) of the organic charge-transfer salt \mbox{$\kappa$-(BEDT-TTF)$_2$Hg(SCN)$_2$Cl} upon slow warming ($q_h\,=\,+1.5\,$K/h) after slow cooling ($q_c\,=\,-3\,$K/h). Red line corresponds to data taken along the out-of-plane $i\,=\,a$ axis, green and blue line to data along the in-plane $i\,=\,b$ and $c$ axis, respectively. The anomalous kinks in $\Delta L_i(T)/L_i$ and the step-like features in $\alpha_i(T)$ at $T_g\,\approx\,$63\,K can be assigned to signatures of the glass-like transition (see main text for a detailed discussion). The jumps in $\Delta L_i/L_i$ and the divergent behavior in $\alpha_i(T)$ at $T_{MI}\,\approx\,30\,$K are related to a charge-order metal-insulator transition the investigation of which is not in the focus of the present study and will be discussed in detail elsewhere \cite{Gati17b}.}
\label{fig:DeltaL}
\end{figure}

Now we will turn to the experimental results of the relative length change $\Delta L_i(T)/L_i$ and the thermal expansion coefficient $\alpha_i(T)\,=\,L_i^{-1}\,\textnormal{d}L_i/\textnormal{d}T$ on \mbox{$\kappa$-(BEDT-TTF)$_2$Hg(SCN)$_2$Cl}. Figure \ref{fig:DeltaL} shows $\Delta L_i(T)/L_i$ (a) and $\alpha_i(T)$ (b) along all three crystallographic axes $i\,=\,a,b,c$ over a wide temperature range $5\,$K$\,\le\,T\,\le\,200\,$K, taken upon warming. At low temperatures $T\,\approx\,30\,$K, $\Delta L_i(T)/L_i$ shows a jump of the length in all crystallographic directions, implying a divergent $\alpha_i(T)$. At this temperature, \mbox{$\kappa$-(BEDT-TTF)$_2$Hg(SCN)$_2$Cl} undergoes a transition from a low-temperature charge-ordered insulating state to a high-temperature metallic state\cite{Yasin12,Drichko14}. Accordingly, we assign this feature to the signature of the metal-insulator transition and label this transition temperature with $T_{MI}$. However, the study of the metal-insulator transition is not in the focus of the present work and will be discussed in detail elsewhere \cite{Gati17b}. Upon warming, a second anomaly can be identified at $T\,\approx\,63\,$K. At this temperature, we observe kink-like anomalies in $\Delta L_i(T)/L_i$ along all three crystallographic axes. These anomalies are reflected by step-like features in $\alpha_i(T)$ (see Fig.\,\ref{fig:DeltaL} (b)). As will be discussed below, the anomalies in $\alpha_i(T)$ can be assigned to glass-like ordering of the EEGs around a characteristic temperature of $T_g\,\approx\,63\,$K. We note that our directional-dependent studies of $\Delta L_i(T)/L_i(T)$ and $\alpha_i(T)$ reveal a strongly anisotropic behavior around $T_g$. The expansivities $\alpha_a$ and $\alpha_b$ are positive below and above $T_g$. This corresponds to an usual increase of the length with increasing $T$, however with an abrupt change of slope around $T_g$ (see Fig.\,\ref{fig:DeltaL} (a)). In contrast, the length along the $c$ axis is increasing with increasing temperature for $T\,<\,T_g$ and decreasing with temperature for $T\,>\,T_g$, giving rise to a sign change of $\alpha_c(T)$ around $T_g$. We stress that the almost null effect in $\Delta L_c/L_c$ along the $c$ axis, observed between $T\,\approx\,80\,$K and 200\,K is consistent with the results of previously published x-ray diffraction data\cite{Drichko14} at room temperature and $T\,=\,100$\,K. The unusual decrease of $\Delta L_c/L_c$ with increasing $T$ observed in the present work is commonly called ``negative thermal expansion'', abbreviated as NTE. It persists up to $T\,\approx\,$140\,K, as revealed by a turning point in $\Delta L_c(T)/L_c$, tantamount to the sign change in $\alpha_c(T)$. A possible origin for this phenomenon in the present salt will be discussed below. First, we present further experimental data on the anomalous contribution to $\alpha_i(T)$ around $T_g\,\approx\,63$\,K.

\begin{figure}
\includegraphics[width=\columnwidth]{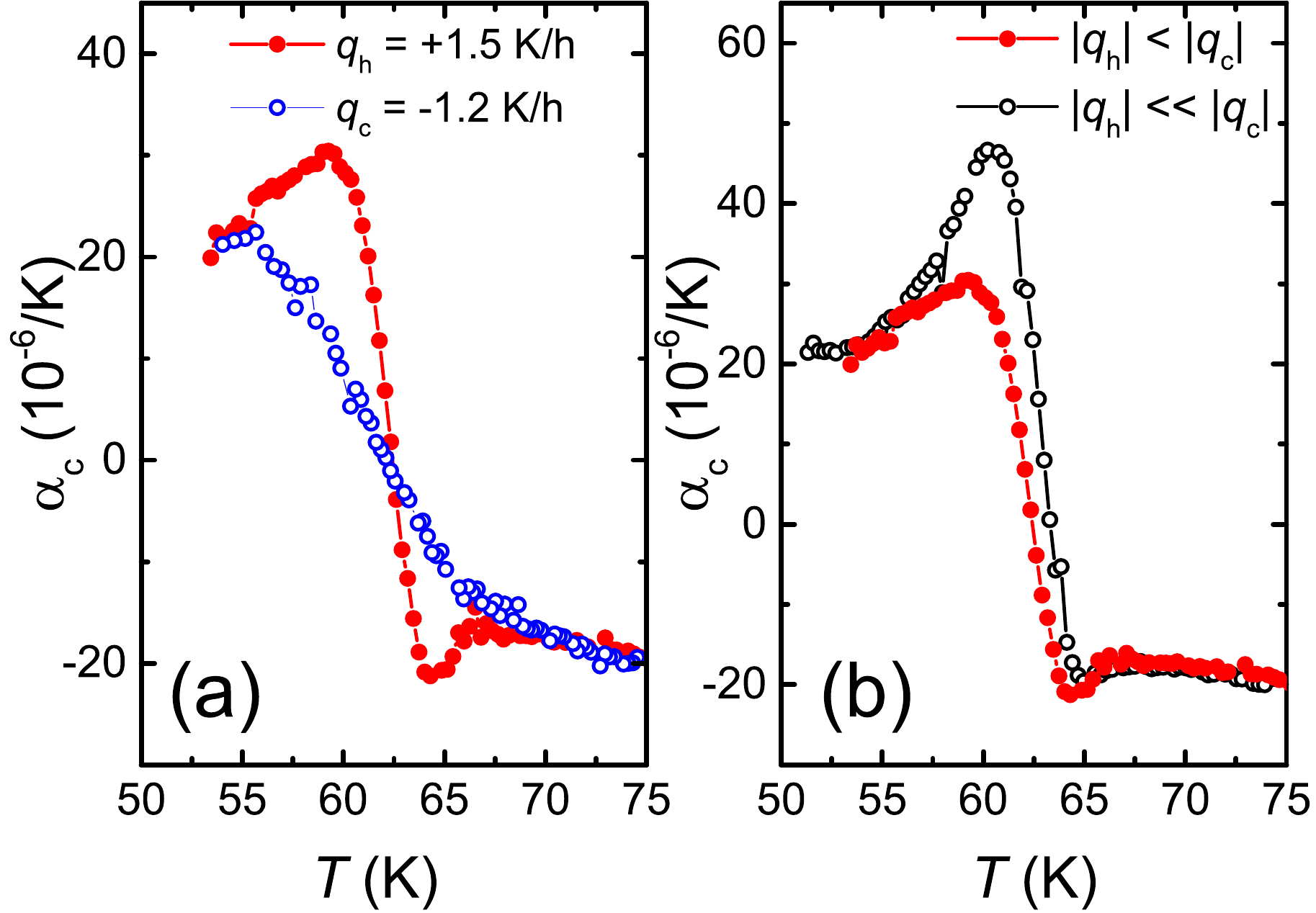} 
\caption{Thermal expansion coefficient $\alpha_c(T)$ of \mbox{$\kappa$-(BEDT-TTF)$_2$Hg(SCN)$_2$Cl} measured along the in-plane $c$ axis around $T_g\,\approx\,63\,$K: (a) Hysteresis between data taken upon slow warming ($q_h\,=\,+1.5\,$K/h, red closed circles) and data taken upon slow cooling ($q_c\,=\,-1.2\,$K/h, blue open circles); (b) Data taken upon slow warming ($q_h\,=\,+1.5\,$K/h) after different cooling procedures: Red closed circles represent data after slow cooling ($q_c\,=\,-$3\,K/h), black open circles represent data taken after fast cooling ($q_c\,=\,-$20.7\,K/h).}
\label{fig:hysteresis}
\end{figure}

Figure \ref{fig:hysteresis} (a) shows a closer look on the thermal expansion coefficient $\alpha_c(T)$, which demonstrates the strongest effect around the glass-like transition at $T_g\,\approx\,63\,$K. This data set was collected upon slowly warming ($q_h\,=\,+1.5\,$K/h after cooldown with $q_c\,=\,-3\,$K/h, red closed circles) as well as upon slowly cooling ($q_c\,=\,-1.2\,$K/h, blue open circles). Upon warming, the step-like feature discussed above is accompanied by characteristic over- and undershoots of $\alpha_c$ at the low- and high-temperature flank of the anomaly, respectively. In contrast, these over- and undershoots are absent upon cooling. Instead, $\alpha_c(T)$ shows a strongly broadened step-like increase. Further away from $T_g$, the data taken upon warming and cooling fall on top of each other. The thermal hysteresis observed here marks an important experimental proof for the glassy nature of the anomaly and rules out a thermodynamic phase transition.

Furthermore, another characteristic aspect for glass-forming systems is the dependence of the expansivity on the cooling $q_c$ and warming $q_h$ rates. This relates on the one hand to (i) the explicit form of the anomaly and on the other hand to (ii) the $T_g(q)$ dependence. Concerning aspect (i), we present in Fig. \ref{fig:hysteresis} (b) two data sets of $\alpha_c(T)$ which were measured upon heating ($q_h\,=\,1.5\,$K/h) after distinctly different cooldown procedures, i.e., after slow cooling with $q_c\,=\,-3\,$K/h (red closed circles) and after comparably fast cooling with $q_c\,=\,-20.7$\,K/h (black open circles). Both data sets reveal a step-like anomaly with over- and undershoot characteristics. Besides the significant shift of the anomaly to higher temperatures upon increasing $|q_c|$ (aspect (ii)) which will be discussed in more detail below the overshoot behavior at the low-temperature side of the anomaly is distinctly larger after fast cooling with $|q_c|\,=\,20.7$\,K/h than after slow cooling with $|q_h|\,\lesssim\,|q_c|\,=\,3$\,K/h.  This is a direct signature of the relaxational phenomena close to $T_g$, which are more pronounced when the glass-forming system is cooled fast $|q_c|\,\gg\,q_h$ and the low-$T$ frozen state is, thus, more disordered.

\begin{figure}
\includegraphics[width=\columnwidth]{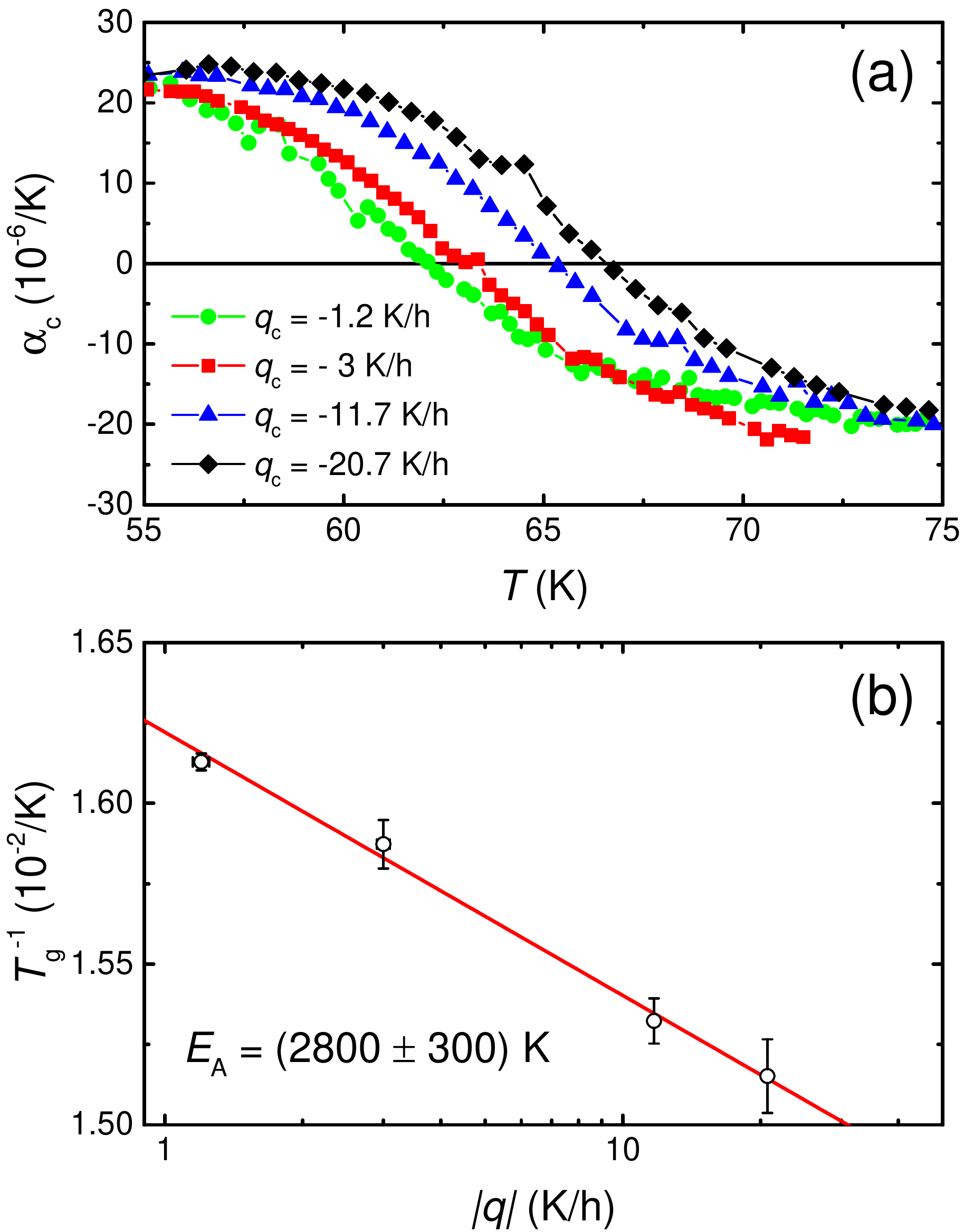} 
\caption{Cooling-rate dependence of the glass-like transition temperature $T_g(|q|)$ and determination of the activation energy $E_A$: (a) Thermal expansion coefficient $\alpha_c(T)$ of \mbox{$\kappa$-(BEDT-TTF)$_2$Hg(SCN)$_2$Cl} measured along the in-plane $c$ axis upon cooling using different cooling rates $-(1.2\,\pm\,0.05)\,$K/h$\,\le\,q_c\,\le\,-(20.7\,\pm\,0.3)\,$K/h; (b) Arrhenius plot of $T_g^{-1}$ vs. $|q|$, yielding an activation energy of $E_A\,=\,(2800\,\pm\,300)\,$K (for details of the analysis, see main text).}
\label{fig:coolingrate}
\end{figure}

Finally, in order to discuss aspect (ii), i.e., the cooling-rate dependence of $T_g$, we present in Fig. \ref{fig:coolingrate} (a) data of $\alpha_c(T)$ upon cooling using different cooling rates in the range -1.2\,K/h\,$\le\,q_c\,\le$\,-20.7\,K/h. It is evident that the anomalous sign change of $\alpha_c$ at $T = T_g$ (see Fig. \ref{fig:hysteresis} (a)), shifts to higher $T$ with increasing $|q_c|$. This is expected for a glass-forming systems in which the relaxation time $\tau$ increases with lowering $T$. To evaluate this shift quantitatively, we determine $T_g$ at a given $q_c$ by using the midpoint of the broad step-like features and include the information of the inverse glass-transition temperature $T_g^{-1}$ vs. $|q|$ in an Arrhenius-like fashion in Fig. \ref{fig:coolingrate} (b). In this representation, the data follow a linear behavior indicating a thermally-activated relaxation time $\tau\,\propto\,\exp(E_A/(k_B T))$. The slope of a linear fit to the present data set can be related to the size of the activation energy barrier $E_A$ in a simple two-level model with thermally-activated relaxation time $\tau$, as outlined in Refs.\,\citen{Mueller02} and \citen{Nagel00}, via
\begin{equation}
\ln|q|\,=\,-\frac{E_A}{k_B T_g}\,+\,\textnormal{const.}
\end{equation}
In the present case, a linear fit (see red line in Fig. \ref{fig:coolingrate} (b)) yields an estimate of the activation barrier energy $E_A\,=\,(2800\,\pm\,300)$\,K. We stress that using this simple Arrhenius-type two-level model for the determination of the activation energy only provides a first estimate of the activation energy. In general, it is known in glass-forming systems that the activation energy can be strongly temperature dependent, in particular close to $T_g$. This temperature dependence is typically described in terms of the so-called Vogel-Fulcher-law. The results of Ref.\,\citen{Mueller15} indeed reveal that the Vogel-Fulcher-law describes the dynamics of the EEGs in other $\kappa$-(BEDT-TTF)$_2X$ salts more accurately in a wide frequency range of at least three orders of magnitude. However, in the present case, only a small frequency range ($4\,\cdot\,10^{-5}$\,Hz\,$\lesssim\,f\,\lesssim\,5\,\cdot\,10^{-4}$\,Hz) is accessible due to the experimentally-limited maximal cooling rate. In this range, the Arrhenius law and the Vogel-Fulcher-law describe the data equally well, and thus the Arrhenius law is suitable for the determination of $E_A$.

\section{Discussion}

The results of the thermal expansion measurements, in particular the step-like contribution to $\alpha_i(T)$ at $T_g\,\approx\,63\,$K, the thermal hysteresis and the strong cooling-rate dependence of $T_g$, provide clear evidence for a glass-like transition in \mbox{$\kappa$-(BEDT-TTF)$_2$Hg(SCN)$_2$Cl}. From the determination of the activation energy $E_A\,=\,(2800\,\pm\,300)\,$K, we can unequivocally assign the glass-like transition to an ordering of the EEGs\cite{Miyagawa95,Akutsu00,Mueller02,Mueller15}. Importantly, we find evidence for \textit{only one} glass-like transition in the present compound despite the existence of two inequivalent EEGs with potential for thermal disorder. In case both EEGs would undergo a glass-like transition they would likely freeze at different temperatures $T_g$, as observed in \mbox{$\kappa$-(BEDT-TTF)$_2$Cu(SCN)$_2$}\cite{Mueller02}. This observation of only one glass-like transition is fully consistent with our predictions from \textit{ab initio} calculations, presented above, in which we identified that only the conformations \textit{S}(1) and \textit{E}(1)  (related to orientational degrees of freedom of the inner EEGs) are close to metastability, but not the conformations \textit{E}(2) and \textit{S}(2). Consequently, we assign the anomalies in $\alpha_i(T)$ to the glassy freezing of the inner EEGs. We note that for other \mbox{$\kappa$-(BEDT-TTF)$_2X$} salts the activation energies $E_A$ determined from thermal expansion measurements were found to be slightly larger than the computed values\cite{Mueller15}. Keeping this in mind, the experimentally determined activation energy of $E_A\,=\,(2800\,\pm\,300)\,$K is fully consistent  with the computed value of $E_A\,=\,$2210\,K for the process \textit{S}(1)$\leftrightarrow$\textit{E}(1) (see more detailed discussion below). We stress that our conclusion can be substantiated by structure determination\cite{Drichko14} down to $T\,=\,100\,$K  which found that the inner EEGs exhibit disorder at this temperature, whereas the outer EEGs are fully ordered. However, we refrain from a comparison with published x-ray diffraction data at $T\,<\,100$\,K, as they were taken at a different instrument than the high-temperature structural data, thereby limiting the possibility of a comparison. \\
Our directional-dependent thermal expansion studies $\alpha_i(T)$ with $i\,=\,a,b,c$ reveal that the freezing of the inner EEGs is accompanied by strongly anisotropic lattice responses, which are particularly pronounced along the out-of-plane $a$ and the in-plane $c$ axis. The anomalous lattice contributions $\Delta \alpha_i(T_g)\,=\,\alpha_i(T\,\rightarrow\,T_g^+)-\alpha_i(T\,\rightarrow\,T_g^-)$ (see Fig. \ref{fig:DeltaL}(b)) amount to $\Delta \alpha_a(T_g)\,=\,+(31\,\pm\,2)\,\cdot\,10^{-6}$/K, $\Delta \alpha_b(T_g)\,=\,+(13\,\pm\,4)\,\cdot\,10^{-6}$/K and $\Delta \alpha_c(T_g)\,=\,-(42\,\pm\,2)\,\cdot\,10^{-6}$/K. From a thermodynamic point of view, the $\Delta \alpha_i(T_g)$ values are related to the change of entropy associated with the EEG freezing $S_{ethy}$ upon application of uniaxial pressure\cite{Mueller02} along the $i$ axis via

\begin{equation}
\frac{\partial S_{ethy}}{\partial p_i} \bigg \vert_{T_g}\,=\,-V_{mol}\,\Delta \alpha_i(T_g).
\end{equation}

By using the molar volume\cite{Drichko14} $V_{mol}\,=\,519\,$cm$^3$, we expect the strongest response of $S_{ethy}$ for uniaxial pressures along the in-plane $c$ axis, for which $\partial S_{ethy}/\partial p_c\,=\,+(2.1\,\pm\,0.1)\,$J/(mol\,K\,kbar) corresponding to an increase of ethylene disorder for increasing uniaxial pressure $p_c$. We note that in a simple two-level model the entropy is $S_{ethy}^{max}\,=\,11.6$\,J/(mol K) at maximum, implying that the entropy is increased by 20\% of $S_{ethy}^{max}$ by a uniaxial pressure of 1\,kbar. This extraordinary pressure sensitivity of $S_{ethy}$ emphasizes the strong metastability of the inner EEG conformation which is suggested by the large $E_A/\Delta E$ obtained in our calculations.

The relative ordering of $|\Delta \alpha_i|$ (with $|\Delta \alpha_c|>|\Delta \alpha_a|>|\Delta \alpha_b|$) is likely related to the directional nature of the short contacts between the inner EEGs and (SCN)$^-$ ligands. The displacement vectors associated with these contacts have components primarily in the $ac$-plane, such that application of pressure along the $a$ or $c$ axis more strongly influences the EEG-anion interactions, and thus the relative stability of the various conformations. This observation strengthens the viewpoint that the coupling to the anions plays an important role in selecting the preferred EEG conformation. 

The negative thermal expansion along the $c$ axis (along the anionic chain direction) that precedes the glass-like transition is also particularly remarkable. A similar effect was also reported for the glass-forming salt \mbox{$\kappa$-(BEDT-TTF)$_2$Cu[N(CN)$_2$]Br}\cite{Wolter07,Souza15}. A plausible scenario is that transverse displacements of the ligands within the $ab$-plane -- away from the chain axis -- cause a shrinkage along the length of the chain\cite{Souza15,Goodwin08,Goodwin05}. This behavior is commonly known as the Poisson effect. We note that the NTE along the $c$ axis is accompanied by an increase in length upon warming along the other two axes, thus indicating that transverse displacements take place. Taken together with the absence of a NTE below $T_g$, the observation of the NTE above $T_g$ supports the notion that the collective EEG-anion motion\cite{Wolter07} freezes out at $T_g$ rather than an individual EEG rotation. This finding is consistent with insights from fluctuation spectroscopy experiments\cite{Mueller15} which revealed non-Arrhenius-like slow dynamics, characteristic for a fragile glass former. It was argued that this implies a significant cooperativity between the EEGs which is most likely mediated by EEG-anion interactions. Indeed, such cooperativity effects might explain the small deviation of the experimentally determined activation energy $E_A\,\approx\,2800$\,K from the calculated one ($E_A\,\approx\,2200$\,K) in the present compound $\kappa$-(BEDT-TTF)$_2$Hg(SCN)$_2$Cl. Whereas the calculations do not take cooperativity effects into account, we determine the activation energy experimentally close to $T_g$ where cooperativity is potentially important and might cause non-Arrhenius-like dynamics. We stress that the present thermodynamic approach does not allow for an analysis of $E_A$ far above $T_g$. However, from the above-mentioned fluctuation spectroscopy experiments it is known that cooperativity tends to increase the activation barrier energy, likely due to the increased number of correlated molecules. In that sense, the larger experimental value of $E_A$ compared to the calculated one suggests that cooperativity among the EEGs, mediated by the anions, leads to a fragile glass-forming state in the present material.

After the detailed analysis of the EEG behavior in \mbox{$\kappa$-(BEDT-TTF)$_2$Hg(SCN)$_2$Cl}, we would like to shortly address why tuning the EEG behavior in this compound could potentially provide important new insights into the physics of dimerized molecular conductors. In a recent study \cite{Gati17b} on the present compound, we provided evidence for electronic ferroelectricity of order-disorder type which originates from charge order within the dimer. We related the occurrence of charge order to a moderate degree of dimerization. Thus, the present material bridges the gap between strongly dimerized materials, often approximated as dimer-Mott systems at 1/2 filling, and non- or weakly dimerized systems at 1/4 filling exhibiting charge order. This conclusion emphasizes the role of the dimerization strength as an important parameter in the field of molecular conductors. Importantly, our present study of the thermal expansion around the glass-like transition suggests that different cooling procedures through $T_g$ can strongly modify the molecular arrangement in the BEDT-TTF plane. Thereby, the intra- and inter-dimer hopping terms, which are the parameters relevant for the electronic structure, are mainly affected. This could potentially be used for tuning the dimerization strength. Likewise, the magnetic frustration ratio, found to be particularly large in the present compound, might be subject to changes induced by different EEG conformations.

\section{Summary}

In conclusion, by employing measurements of the thermal expansion, we provide clear evidence for a glass-like ordering transition at $T_g\,\approx\,63\,$K in the organic charge-transfer salt \mbox{$\kappa$-(BEDT-TTF)$_2$Hg(SCN)$_2$Cl}. Similar to other \mbox{$\kappa$-(BEDT-TTF)$_2X$} salts, orientational degrees of freedom of the ethylene endgroups (EEGs) of the BEDT-TTF molecule were identified to be responsible for the glassy behavior. In this regard, the present salt \mbox{$\kappa$-(BEDT-TTF)$_2$Hg(SCN)$_2$Cl} is special as the two inequivalent EEGs behave distinctly different, as one of them orders smoothly, whereas the other one freezes in a glassy manner. This result is consistent with \textit{ab initio} calculations, which estimate energy differences $\Delta E$ and activation energies $E_A$ for the different conformations. Distinctly different interactions between the inequivalent EEGs and the anions lead to one metastable state, as evidenced by a large ratio $E_A/\Delta E$, and one strongly confined state with small $E_A/\Delta E$. Thus, our results confirm the concept proposed in Ref.~\citen{Mueller15} that $E_A/\Delta E$ is a suitable parameter to quantify the tendency of a system towards glass-like ordering. The identification of the peculiar EEG ordering in \mbox{$\kappa$-(BEDT-TTF)$_2$Hg(SCN)$_2$Cl} confirms that the interaction between the EEGs and the anions is the decisive factor for the occurrence of glass-like freezing in the \mbox{$\kappa$-(BEDT-TTF)$_2X$} organic charge-transfer salts. As the EEG vibrational degrees of freedom are known to couple strongly to the electronic degrees of freedom in this material class, it is interesting to investigate in the future how ground state properties of the strongly correlated electron system in \mbox{$\kappa$-(BEDT-TTF)$_2$Hg(SCN)$_2$Cl} are influenced by the presence of two EEG subsystems with distinctly different temperature-dependent vibrational properties.

\begin{acknowledgments}
We acknowledge useful discussions with Benedikt Hartmann. Research performed in Frankfurt was supported by the German Science Foundation via the Transregional Collaborative Research Center SFB/TR49: Condensed Matter Systems with Variable Many-Body Interactions. SMW thanks NSERC Canada for a postdoctoral fellowship. JAS acknowledges support from the Independent Research and Development program while serving at the NSF.
\end{acknowledgments}

\bibliographystyle{apsrev}
\bibliography{Lit}

\begin{thebibliography}{62}
\bibitem[{fus()}]{fussnote1}
\bibinfo{note}{We note that a step-like increase in $C(T)$ is related to a
  step-like change of $\alpha_i(T)$ via the Gr\"uneisen relation. However, the
  sign of the anomalous contribution to $\alpha_i(T)$ is given by the
  Gr\"uneisen parameter, i.e., by the pressure dependence of the corresponding
  energy scale.}
\end{thebibliography}

\clearpage


\section*{Supplementary material}

# Supplementary Information: Insights from experiment and *ab initio* calculations into the glass-like transition in the molecular conductor $\kappa$-(BEDT-TTF)$_2$Hg(SCN)$_2$Cl

Elena Gati, Stephen M. Winter, John A. Schlueter, Harald Schubert, Jens Müller, and Michael Lang

(Dated: January 28, 2018)

## I. *AB INITIO* RESULTS: CONFORMATION ENERGIES

In the main text, we present calculated energy differences between different ethylene endgroup (EEG) conformations. Here, we provide further details on the determination of these values.

In $\kappa$-(BEDT-TTF)$_2$Hg(SCN)$_2$Cl, the four EEGs within a given dimer may each adopt two possible orientations, leading to a total of $2^4 = 16$ possible conformations. Of these, there are four conformations in which both molecules adopt a staggered orientation, four in which both molecules adopt an eclipsed orientation, and eight conformations representing a mixture of staggered and eclipsed orientation. In order to label the different conformations, we considered that each molecule within the dimer may adopt four possible conformations, two of which are staggered (labeled $S(1)$ and $S(2)$), and two of which are eclipsed (labeled $E(1)$ and $E(2)$). As described in detail in the main text, $E(1)$ is related to $S(1)$ via the flipping of an inner EEG (C7/C8), while $E(2)$ is related to $S(1)$ via the flipping of an outer EEG (C9/C10). Finally, $S(2)$ is related to $S(1)$ via the flipping of both EEGs (C7/C8 and C9/C10) within the same molecule.

We studied the energetics of all 16 conformations within one dimer. The results are shown in the table below. The first two rows indicate the conformations of the two molecules within each dimer. The third row shows the *ab initio* energies (rounded to the nearest 100 K) for each conformation obtained after relaxing the positions of all EEGs. The final row shows the experimental population of the 16 different conformations at room temperature, as estimated from the refined occupancies from the x-ray structure[1].

In the manuscript, we considered the energy cost $\Delta E_i$ to convert one molecule from $S(1)$ to the $E(1)$, $E(2)$, and $S(2)$ conformation. In order to obtain these values, we averaged all energy contributions of changing either an inner or an outer EEG according to the experimental room-

| molecule 1 | $S(1)$ | $S(1)$ | $E(1)$ | $E(1)$ | $S(1)$ | $E(2)$ | $S(1)$ | $S(2)$ | $E(1)$ | $E(2)$ | $E(1)$ | $S(2)$ | $E(2)$ | $E(2)$ | $S(2)$ | $S(2)$ |
|---|---|---|---|---|---|---|---|---|---|---|---|---|---|---|---|---|
| molecule 2 | $S(1)$ | $E(1)$ | $S(1)$ | $E(1)$ | $E(2)$ | $S(1)$ | $S(2)$ | $S(1)$ | $E(2)$ | $E(1)$ | $S(2)$ | $E(1)$ | $E(2)$ | $S(2)$ | $E(2)$ | $S(2)$ |
| $E(10^2\,\mathrm{K})$ | 0 | | 3 | | 7 | | 15 | | 19 | | 20 | | 24 | | 30 | | 36 | | 42 |
| population (%) | 19 | | 15 | | 12 | | 5 | | 4 | | 4 | | 3 | | 2 | | 1 | | <1 |

TABLE I. List of computed energies $E$ and experimentally-determined population[1] of the different conformations of the EEGs within one dimer. $S(1)$ and $S(2)$ denote staggered conformations and $E(1)$ and $E(2)$ denote eclipsed conformations (see main text for details).

temperature occupancies.

---



\end{document}